\documentclass[natbib,apjl]{emulateapj}
\usepackage{epsfig}
\usepackage{graphicx}
\usepackage{epstopdf}
\usepackage{url}

\newcommand{\actaa}{Acta Astronomica}
\newcommand{\icarus}{Icarus}
\newcommand{\av}{A$_V$}

\newcommand{\lx}{L$_{X}$}
\newcommand{\msun}{M$_{\odot}$}
\newcommand{\logl}{log(L/L$_{\odot}$)}
\newcommand{\ergs}{erg\,s$^{-1}$}

\newcommand{\loglxlbol}{log($L_{\rm X}$/$L_{bol}$)}

\newcommand{\kms}{km\,s$^{-1}$}
\newcommand{\masyr}{mas\,yr$^{-1}$}
\newcommand{\mv}{M$_{\rm V}$}

\newcommand{\mmoon}{M$_{Moon}$}

\slugcomment{Accepted for publication in Astronomical Journal 2012 January 2}
\shorttitle{Disk and Ring Eclipses}
\shortauthors{Mamajek et al.}

\begin{document}

\title{PLANETARY CONSTRUCTION ZONES IN OCCULTATION:\\
  DISCOVERY OF AN
  EXTRASOLAR RING SYSTEM TRANSITING A YOUNG SUN-LIKE STAR\\
  AND FUTURE PROSPECTS FOR DETECTING ECLIPSES BY CIRCUMSECONDARY AND
  CIRCUMPLANETARY DISKS}

\author{
Eric E. Mamajek\altaffilmark{1,2}, 
Alice C. Quillen\altaffilmark{1}, 
Mark J. Pecaut\altaffilmark{1}, 
Fred Moolekamp\altaffilmark{1}, 
Erin L. Scott\altaffilmark{1},\\
Matthew A. Kenworthy\altaffilmark{3}, 
Andrew Collier Cameron\altaffilmark{4} \& 
Neil R. Parley\altaffilmark{4}}

\altaffiltext{1}{Department of Physics and Astronomy, University of
  Rochester, Rochester, NY 14627-0171}
\altaffiltext{2}{Current address: Cerro Tololo Inter-American Observatory, 
Casilla 603, La Serena, Chile}
\altaffiltext{3}{Leiden Observatory, Leiden University, P.O. Box 9513, 2300 RA Leiden, The Netherlands}
\altaffiltext{4}{School of Physics and Astronomy, University of St Andrews,
  North Haugh, St Andrews, Fife KY16 9SS}

\begin{abstract}
  The large relative sizes of circumstellar and circumplanetary disks
  imply that they might be seen in eclipse in stellar light curves.
  We estimate that a survey of $\sim$10$^4$ young ($\sim$10 million
  year old) post-accretion pre-main sequence stars monitored for
  $\sim$10 years should yield at least a few deep eclipses from
  circumplanetary disks and disks surrounding low mass companion
  stars.  We present photometric and spectroscopic data for a pre-main
  sequence K5 star (1SWASP J140747.93-394542.6 = ASAS J140748-3945.7),
  a newly discovered $\sim$0.9 M$_{\odot}$ member of the $\sim$16
  Myr-old Upper Centaurus-Lupus subgroup of Sco-Cen at a kinematic
  distance of 128\,$\pm$\,13 pc.  This star exhibited a remarkably
  long, deep, and complex eclipse event centered on 29 April 2007 (as
  discovered in SuperWASP photometry, and with portions of the dimming
  confirmed by ASAS data). At least 5 multi-day dimming events of
  $>$0.5 mag are identified, with a $>$3.3 mag deep eclipse bracketed
  by two pairs of $\sim$1 mag eclipses symmetrically occurring $\pm$12
  days and $\pm$26 days before and after. Hence, significant dimming
  of the star was taking place on and off over at least a $\sim$54 day
  period in 2007, and a strong $>$1 mag dimming event occurring over a
  $\sim$12 day span. We place a firm lower limit on the period of 850
  days (i.e. the orbital radius of the eclipser must be $>$1.7 AU and
  orbital velocity must be $<$22 km/s).  The shape of the light curve
  is similar to the lop-sided eclipses of the Be star EE Cep. We
  suspect that this new star is being eclipsed by a low-mass object
  orbited by a dense inner disk, further girded by at least 3 dusty
  rings of optical depths near unity. Between these rings are at least
  two annuli of near-zero optical depth (i.e.  gaps), possibly cleared
  out by planets or moons, depending on the nature of the secondary.
  For possible periods in the range 2.33-200 yr, the estimated total
  ring mass is $\sim$8-0.4 \mmoon (if the rings have optical opacity
  similar to Saturn's rings), and the edge of the outermost detected
  ring has orbital radius $\sim$0.4-0.09 AU. In the new era of
  time-domain astronomy opened by surveys like SuperWASP, ASAS, etc.,
  and soon to be revolutionized by LSST, discovering and
  characterizing eclipses by circumplanetary and circumsecondary disks
  will provide us with observational constraints on the conditions
  which spawn satellite systems around gas giant planets and planetary
  systems around stars.
\end{abstract}


\section{Introduction}

The radii of circumstellar and circumplanetary disks can vastly exceed
those of stars and planets.  A companion star of a young stellar
binary system can host a circumstellar disk, and likewise a giant
planet in a young stellar system can host a circumplanetary disk.
Because the disks are large, the probability that a randomly oriented
system exhibits eclipses may not be negligible.  These disks are
particularly interesting as they could be seen in eclipse during the
epoch of planet formation \citep[in the case of a companion
circumstellar disk; ][]{Galan10} or during the epoch of satellite
formation (in the case of a circumplanetary disk).  In this paper we
consider the possibility of discovering eclipses by dust disks of
low-mass companions in long period orbits.  With the advent of long
term and large scale photometric surveys, strategies can be developed
to discover young systems eclipsed by disks.
 
Some well-known long period eclipsing systems have been interpreted in
terms of occulting dark disks associated with an orbiting companion,
with the best examples being $\epsilon$ Aurigae \citep{Guinan02,
  Kloppenborg10, Chadima11}, EE Cep \citep{Mikolajewski99, Graczyk03,
  Mikolajewski05, Galan10}, and the newly identified
OGLE-LMC-ECL-17782 \citep{Graczyk11}.  EE Cep exhibits long (30-90
day) asymmetric eclipses with a period of 5.6 years and depth of
$\sim$0.6 -- 2.1 magnitudes.  The primary object is a B5e giant star
and only primary eclipses are seen \citep{Mikolajewski99}. Structure
seen in the wings of the eclipse has recently been interpreted in
terms of rings and gaps in a forming planetary system around a lower
mass secondary \citep{Galan10}.  $\epsilon$ Aurigae is the eclipsing
system with the longest known period of 27.1 years.  Only primary
eclipses are seen and they last almost 2 years.  While the mass of the
object hosting the dark occulting disk exceeds that of visible F star,
the masses of the two stars are not well constrained. However,
emission lines and UV emission suggest that the hidden object is a
B-type star \citep[see discussion by ][]{Chadima11}.  Infrared
emission from the disk was detected with IRAS \citep{Backman85}.
OGLE-LMC-ECL-17782 \citep[MACHO J053036.7-690625; ][]{Graczyk11}, is a
13-day eclipsing binary in the Large Magellanic Cloud that
demonstrates wide, flat-bottomed eclipses like $\epsilon$ Aur, but
there are changes in the light curve from eclipse to eclipse, and
transient features visible at other phases. \citet{Graczyk11} suggest
that the system is a detached binary\footnote{The author dereddened
  the UBV photometry for OGLE-LMC-ECL-17782 from \citet[][ catalogued
  as M2002 \#148104]{Massey02} using the Q-method, and finds that the
  primary is most likely a slightly evolved $\sim$B2 star with
  (B-V)$_o$ $\simeq$ -0.23 and M$_V$ $\simeq$ -3.7. However this
  calculation does not take into account the difference in
  metallicities between LMC and local Galactic massive stars.} where
the secondary is ``{\it partially hidden within a semi-transparent,
  dark, elongated body or disk}'' and there are likely ``{\it
  transient structures in the system (disk debris?)  responsible for
  additional minima at different orbital phases when one of the stars
  is hidden behind them}.''

In this paper we present the discovery of a solar mass pre-main
sequence non-accreting star exhibiting a long, unusual eclipse similar
to those seen in EE Cep and $\epsilon$ Aurigae. The mass of the star
and lack of detected infrared emission suggest that the host object
for the eclipsing disk is low mass, possibly a low mass star, brown
dwarf, or giant planet.  Hence we consider both {\it circumsecondary}
and {\it circumplanetary disks} as possible occulting objects.

Why should one also consider {\it circumplanetary disks} associated
with giant planets?  First, natural satellites are a ubiquitous
feature among the giant planets in our solar system, and most likely
among extrasolar gas giants. The existence of such satellites, and the
H/He-rich atmospheres of gas giants hint these planets likely formed
with large gas and dust disks ($r_{planet} << r_{disk} < r_{Hill}$)
that were originally accretion disks feeding from circumstellar
material. These disks would then evolve passively after the
circumstellar reservoir was depleted (Sec. 2), with matter accreting
on to the planet, grain growth and proto-satellite accretion, and
depletion through other mechanisms (e.g.  Poynting-Robertson drag,
radiation pressure, photoionization, etc.).

A simple thought experiment illustrates the potential observability of
moon-forming circumplanetary disks around young gas giants (and indeed
this was the back of the envelope calculation that spawned our
interest in the interpretation of the eclipsing star discussed in Sec.
3). If one were take the Galilean satellites of Jupiter and grind them
up into dust grains, and spread the grains uniformly between Jupiter
and Callisto's orbit, one would have a dusty disk of optical depth
$O$(10$^{5}$). The size of such a proto-moon disk in this case would
be a few solar radii -- i.e. large enough and optically thick enough
to potential eclipse a star's light. Of course such a disk need not be
face on -- more likely the disk would have a non-zero inclination
respect to the planet-star orbital plane, so the star need not be
completely geometrically eclipsed by such a circumplanetary disk.  The
rings of Saturn have optical depth near $\sim$1 even at a relatively
old age (4.6 Gyr), however the vast majority of mass orbiting Saturn
is locked up in satellites ($M_{rings}$ $\simeq$ 10$^{-4}$
$M_{satellites}$).  Presumably a disk of much higher optical depth and
significant radial substructure existed during the epoch of satellite
formation. While there have been studies investigating the
detectability of thin, discrete planetary rings similar to Saturn's
\citep[e.g.][]{Barnes04, Ohta09}, there has been negligible
investigation of the observability of the dense proto-satellite disks
that likely existed during the first $\sim$10$^7$ years.  Relaxing the
assumptions about the size, mass, composition and structure of the
disk in our back-of-the-envelope calculation has little impact on the
feasibility of the idea that {\it dusty disks of high optical depth
  may be a common feature of young gas giant planets, and such objects
  may be observable via deep eclipses of young stars}.

We first estimate the timescale of an eclipse by a circumsecondary or
circumplanetary disk in Sec. 2.  This is done first so that the
photometric light curve of our object can be interpreted in terms of
eclipse models.  We then present and discuss the properties of our
candidate long period eclipsing system in Sec. 3.  In Sec.  4 we
discuss the probability of detecting eclipses using time series
photometry of large samples of young stars.  A discussion and summary
follows in Sec. 5.
 
\section{Circumsecondary and circumplanetary Disk Eclipses}

The multiplicity of class I young stellar objects (YSOs) in embedded
clusters and pre-MS stars in young associations is high and ranges
from 20 to 60\% for the $\sim$10$^{1.5}$-10$^{2.5}$ AU separation
range and mass ratios of $\sim$0.1-1 (see the comprehensive review by
\citealt{Duchene07} and also \citealt{Kraus11}).  Among the binaries
found in young clusters, stars in different stages of disk evolution
are not rare \citep{Hartigan03, McCabe06, Monin07, Prato10}.  These
binaries, known as ``mixed pairs'', have one star hosting a disk or
actively accreting and the other lacking a disk or signatures of
accretion.  While the two most well known long period eclipsing disk
systems (EE Cep and $\epsilon$ Aurigae) involve massive stars, the
large fraction of binary systems in young associations and clusters
suggest that long period eclipsing circumsecondary disk systems may be
discovered in lower mass systems.

Recent explorations of circumplanetary disks separate the disk
evolution into two phases \citep{Alibert05, Ward10}.  In the first
phase the circumplanetary disk is fed by material from the
circumstellar disk and the circumplanetary disk acts like an accretion
disk.  In the second phase the circumstellar disk has dissipated, and
the viscosity of the circumplanetary disk drives both accretion onto
the planet and causes the disk to spread outwards.  The lack of
differentiation of Callisto \citep{Anderson01} suggests that the
accretion or formation timescale for all the Galilean satellites was
prolonged \citep{Canup02} and would have happened during the second
phase of evolution after the circumstellar disk dissipated
\citep{Canup02, Alibert05, Ward10, Mosqueira10}. Also, modeling of
Iapetus, the outermost regular satellite of Saturn, suggests that it
formed 3--5 Myr after the formation of Ca-Al inclusions
\citep{CastilloRogez09}.  Since Iapetus survived Type I migration it
must have formed near the end of substantial accretion onto Saturn
from the circumsolar nebula.  \citet{Mosqueira10} suggest that
Iapetus' large separation from Saturn's principal satellite Titan is
suggestive that the remnant circum-Saturnian nebula may have had two
components: a dense inner disk that spawned most of Saturn's regular
satellites, out to the centrifugal radius near Titan, and a disk of
much lower density beyond the centrifugal radius out -- perhaps to
Phoebe's orbit. Because of the extended estimated circumplanetary disk
lifetime, we can consider the possibility that circumplanetary disks
can be seen in eclipse against a young central star after the
dissipation of the circumstellar disk.

We consider two bodies with masses $m_1, m_2$ in a circular orbit with
semi-major axis $a_B$ and mass ratio $\mu \equiv {m_2 \over m_1 +
  m_2}$.  In the case of a stellar binary the more massive star is
$m_1$ and the secondary is $m_2$.  In the case of a system with a
single planet, $m_2$ is the mass of the planet and $m_1$ the mass of
the central star.  The masses and semi-major axis set the Hill or
tidal radius of the secondary $r_H \equiv a_B \left({\mu\over
    3}\right)^{1/3}$.

A disk surrounding $m_2$ is described with two parameters, the radius
at which its optical depth is of order unity, $r_d$, and the obliquity
or axial tilt of the disk system, $\epsilon$ with respect to the axis
defining the orbital plane.  The angle $\epsilon$ is zero for a disk
that lies in the orbital plane.  It is convenient to define a size
ratio $\xi \equiv r_d/r_H$ that represents the size of the disk in
Hill or tidal radii.

Studies of giant planets have made estimates for the size ratio $\xi$.
Based on a centripetal radius argument \citet{Quillen98} estimated
that an accreting circumplanetary disk would have $\xi = 1/3$.
Hydrodynamic simulations of planets embedded in circumstellar disks
also can find $\xi \sim 0.3$ \citep{Ayliffe09}.  A tidal truncation
argument suggests $\xi \sim 0.4$ \citep{Martin11}.  Theoretical models
for Jupiter's circumplanetary disk accounting for Galilean satellite
formation after the dissipation of the circumstellar disk estimate
smaller radii of $\xi \sim$ 0.1 -- 0.2 \citep{Canup02, Magni04,
  Ward10}.  For reference, the outermost of the Galilean satellites,
Callisto, currently has a semi-major axis that is only a small
fraction of Jupiter's Hill radius $a\approx 0.0355 r_H$.  Jupiter's
Hill radius is about $743 R_J$ where $R_J$ is the radius of Jupiter.
Estimates of the Hill radii $r_H$ and the ratio of the orbital radii
to Hill radii ($r_{sat}$/$r_H$) for the outermost large, regular
satellites for the giant planets in our solar system are summarized in
Table \ref{hillradii}. Column 6 of Table \ref{hillradii} estimates the
timescale for a circumplanetary disk with outer radius equal to the
orbital radius of the outermost regular satellite to transit in front
of the Sun, given the planet's mean orbital velocity.

The results suggest that substantial circumplanetary disks of
sufficient surface density for satellites to accrete must have existed
around the giant planets in our solar system for some period during
the post-T Tauri phase for our Sun.  These circumplanetary disks
likely had outer radii of $\xi$ $>$ 0.01-0.05, and could have been
detectable by eclipses with $\sim$1-10 day timescales to observers
along opportune lines-of-sight.

\begin{deluxetable}{lllllrlrcrrrr}
  \setlength{\tabcolsep}{0.03in} \tablewidth{0pt}
  \tablecaption{Hill Radii and Orbital Radii of Outermost Regular Satellites for Large 
Planets \label{hillradii}} \tablehead{
    {(1)}   &{(2)}     &{(3)}      &{(4)}      &{(5)}            &{(6)}\\
    {Planet}&{R$_{h}$} &{$v_{orb}$}&{Outermost}&{$r_{sat}$/$r_H$}&{t$_{transit}$}\\
    {}      &{(AU)}    &{\kms}     & {Reg. Sat.}&{}              &{(days)}}
\startdata
Jupiter & 0.36 & 13.1 & Callisto &  0.035 & 3.3\\
Saturn  & 0.44 &  9.7 & Iapetus  &  0.054 & 8.5\\
Uranus  & 0.47 &  6.8 & Oberon   &  0.008 & 2.0\\
Neptune & 0.78 &  5.4 & ...      &  ...   & ...\\
\enddata
\tablecomments{Column (2) is the Hill radius $r_H$ in AU, (3) is the
  planet's mean orbital velocity in \kms, (4) is the name of the
  outermost large regular satellite for each planet, (5)
  $r_{sat}$/$r_H$ is the ratio of that moons orbital radius to the
  Hill radius, and (6) is the timescale $t_{transit}$ that a
  circumplanetary disk with outer radius equal to the orbital radius
  of the outermost large regular satellite to transit in front of the
  Sun, given the planet's orbital velocity.  Neptune does not have a
  system of large regular satellites. Its principal moon Triton is in
  a retrograde orbit, and was likely captured as a component of a
  binary dwarf planet \citep{Agnor06}}
\end{deluxetable}

A disk in a binary system might extend all the way to its Roche radius
\citep{Lin93, Artymowicz94, Andrews10} with $\xi \sim 1$.  For
example, the disk of HD141569A \citep{Clampin03} may extend to its
Roche radius when the secondary is at pericenter and approaches
HD141569A \citep{Augereau04, Quillen05}.  The truncated disks of HD
98800 and Hen 3-600 are consistent with tidal truncation
\citep{Andrews10}.
 
A circular orbit for $m_2$ would have a circumference of $\sim 2 \pi
a_B$.  The disk extends a distance $\pm r_d \sin \epsilon$ above and
below the orbital plane.  Thus the area of a cylinder that could be
intersected by an eclipsing line of sight is $A \sim 4 \pi a_B r_d
\sin \epsilon$. Here we have neglected the thickness of the disk at
low obliquity.  If the disk is seen edge on then $A \sim 4 \pi a_p
h_d$ where $h_d$ is the scale height of the disk near its opacity
edge.  We can define a factor

\begin{equation}
y(\epsilon) = \max \left( {h_d \over r_d}, \sin \epsilon \right).
\end{equation}  

To estimate the probability that a system containing a disk is
oriented so that it would exhibit eclipses we divide this area by $4
\pi a_p^2$

\begin{equation}
p_{orient} \sim {r_d y(\epsilon) \over a_B} \sim \xi \mu^{1/3} 3^{-1/3} y(\epsilon)
\label{eqn:p}
\end{equation} 

This probability is independent of the binary or planet's semi-major
axis.

Studies of primordial circumplanetary disks have calculated their
thermal structure \citep{Canup02, Alibert05}.  The circumplanetary
disk aspect ratio $h/r$ is predicted to be in the range 0.1 - 0.3
where $h$ is the vertical scale height a radial distance $r$ from the
planet's center \citep[e.g., see Figure 9 ;][]{Alibert05}.
Hydrostatic equilibrium relates the temperature $T$ of a gaseous disk
to its vertical scale height and the speed $\Omega$ of objects in
orbit ($h$ $\sim$ $c_s$/$\Omega$, where sound speed $c_s$ =
($kT$)$^{1/2}$($\mu$m$_H$)$^{-1/2}$), $k$ is Boltzmann's constant,
$\mu$ is the mean molecular weight, and $m_H$ is the mass of
hydrogen).  Due to the low circular velocities for objects in orbit
about a planet, a gaseous circumplanetary disk would not have a small
aspect ratio until its gas has dissipated.

The temperature and scale height of a circumstellar disk is set from
the dominant source of heat, which is from absorption of stellar
radiation unless the accretion rate is high.  A good rough guideline
covering a wide range of conditions is aspect ratio $h/r \sim 0.05$ --
0.1 \citep[e.g., see Figure 1 by ][]{Edgar07}.  The factor
$y(\epsilon)$ also depends on the orientation of the disk.  Strongly
misaligned disks should be brought into rough alignment by shocks
associated with tidal torques in only about 20 binary orbital periods
\citep{Bate00}.  For modest misalignment angles, misalignment can
persist over considerably longer periods.  Disks have been imaged in
wide binary T Tauri stars that are misaligned with the binary orbital
plane \citep[e.g., HK Tau; ][]{Stapelfeldt98}.  If the system is not a
binary but a triplet or a quadruple system then misalignment may be
more common \citep[as discussed by ][]{Prato10}.  \citet{Prato10}
emphasize (see their section 4) that ``even binaries with separations
of a few tens of AU -- or less -- cannot be assumed to harbor aligned
disks coplanar with binary orbits.''  Hence the obliquity distribution
for circumsecondary disks may be wide.

Evaluating eqn. \ref{eqn:p} with illustrative parameters we estimate
that the fraction of randomly oriented systems (at a given mass ratio)
hosting a disk that are oriented so they could exhibit an eclipse is

\begin{equation}
f_{orient} \sim 0.004 \left({\xi \over 0.2} \right) \left({m_2 \over m_J}\right)^{1\over 3} 
\left({m_1 + m_2\over M_\odot}\right)^{-{1\over 3}} 
\left({{\bar y} \over 0.3} \right).
\label{eqn:fo}
\end{equation}

Here we have taken $\bar y$ to be the mean of the distribution of
$y(\epsilon)$ that depends on the obliquity and disk height
distribution and $M_J$ is the mass of Jupiter.  Because of the large
size of a circumplanetary or circumsecondary disk the fraction
$f_{orient}$ of objects capable of giving eclipses is not low, and
that such eclipses will be seen.

The timescale for the eclipse to occur will depend on the angular
rotation rate of the orbit

\begin{eqnarray}
t_{eclipse} &\sim& \sqrt{a_B^3 \over G(m_1+ m_2)} {2 r_d \over a_B} \sim P \xi \mu^{1/3} \pi^{-1} 3^{-1/3} \\
&\sim & 1.5 {\rm ~days} \left({\xi \over 0.2} \right) \left({m_2 \over m_J}\right)^{1\over 3} \left({m_1+m_2 \over M_\odot}\right)^{-{1\over 3}} \left({a_B \over 1 {\rm AU} } \right)^{3\over 2}\nonumber 
\label{eqn:days}
\end{eqnarray}

where we have assumed that the radius of the occulted object $R_1 \ll
r_{d}$ and $P$ is the orbital period of the planet or binary.

The fraction of the orbit spent in eclipse is 

\begin{equation}
f_{eclipse} = {t_{eclipse} \over P } \sim \xi \mu^{1/3} \pi^{-1} 3^{-1/3},
\label{eqn:fe}
\end{equation}

and is independent of the semi-major axis.

Once a disk eclipse candidate is identified one could search for
reflected star light from the disk.  The area intersecting light from
the star $A \sim 4\pi r_d^2 \sin y(\epsilon) c$ where the order unity
factor $c$ depends on the orientation of the disk.  The fraction of
reflected star light would be

\begin{eqnarray}
f_r 	\sim \left({r_d \over a_p}\right)^2 y(\epsilon) c' 
	\sim \xi^2 \mu^{2/3} y(\epsilon) c'  
\end{eqnarray}

where the order unity factor $c'$ depends on $c$, the disk's albedo
and the dependence on scattering angle.  The difference in magnitude
between the reflected light of the disk and

\begin{eqnarray}
\delta m \sim  9.8 - 2.5 \log_{10} \left[ \left({\xi \over 0.2}\right)^2 \left({\mu \over 10^{-3}} \right)^{2\over 3} \left({y(\epsilon) c' \over 0.3} \right)  \right].
\end{eqnarray}

This level of magnitude difference is not extremely high suggesting
that it may be feasible to detect reflected light from circumplanetary
disks with an adaptive optics system.

\subsection{Disk Lifetimes}

For young binary systems in nearby star-forming regions (with typical
ages of $<$3 Myr) the estimated fraction of mixed systems with primary
a weak lined T Tauri star and the secondary a classical T Tauri star
is not low, and could be as large as $\sim 1/3$ \citep{Monin07}.  The
mixed systems imply that the circumstellar disks around the primary
and secondary can have lifetimes that differ by a factor of about 2
\citep{Monin07}.  A tidally truncated disk around a low mass secondary
is expected to have a shorter accretion lifetime than the primary's
disk \citep{Armitage99}, however there are other environmental factors
such as dispersal of the host molecular cloud, birth cluster density,
and disk evaporation that can influence multiplicity and disperse
disks \citep{Monin07, Prato10}.

We can estimate the lifetime of a circumplanetary disk by scaling from
models for the proto-Jovian disk.  The orbital period of a particle in
orbit about a planet with radius near the Hill radius is approximately
that of the planet in orbit about the star.  The lifetime of a
circumplanetary disk likely depends on the orbital period at its outer
edge and so depends on $\xi^{3/2}$ times the planet's orbital period.
If $\xi$ is similar for different circumplanetary disks, the lifetime
of the secondary phase of these disk (after circumstellar disk
dissipation) should be proportional to the planet's semi-major axis to
the 3/2 power. We can use this scaling relation to estimate the
lifetime of circumplanetary disks at larger distances from the star
than Jupiter.  The lifetime of the second phase of Jupiter's
circumstellar disk is estimated to be of order a million years
\citep{Canup02, Alibert05, Ward10}.  The lifetime of a circumplanetary
disk at 25 AU we estimate could be about 10 times longer (or of order
$10^7$ years) and at 100 AU (such as Fomalhaut b) a 100 times longer
or of order $10^8$ years.

The planet Fomalhaut b has been detected in two visible bands only
\citep{Kalas08}.  While the detected object is in orbit about
Fomalhaut it's color is consistent with that of reflected light from
the star.  These observations led Kalas et al. to suggest that the
planet hosts a circumplanetary ring akin to Saturn's \citep[see
][]{Arnold04}, which would extend to at least 20 Jupiter radii for an
assumed albedo of 0.4 to recover the observed fluxes.  Another
possibility is that this is reflected light from a gaseous
circumplanetary disk.  Due to the large semi-major axis of the planet,
(and so large Hill radius) the lifetime of this disk would exceed that
estimated for Jupiter's circumplanetary disk, making this possibility
more tractable.  The light from Fomalhaut b is unresolved so the
emitting object must be confined to a region smaller than the Hubble
Space Telescope Advanced Camera point spread function (PSF) full width
at half-maximum (FWHM) of 0.5 AU \citep{Kennedy11}.  The ratio of 0.5
AU to the planet's semi-major axis of 119~AU gives a constraint $\xi^3
\mu = 2.2 \times 10^{-7}$.  For $\xi=0.2$ this gives a planet mass
ratio of $2 \times 10^{-5}$ and for $\xi = 0.1$ of $2 \times 10^{-4}$
so planet masses ranging from Saturn to Neptune mass and lower than
estimated by \citet{Chiang09} but consistent with that predicted by
\citet{Quillen06}.

\section{A Candidate Eclipsing Disk}

\subsection{The Star}

We are currently conducting a large scale spectroscopic survey for new
low-mass members of the Sco-Cen OB association (Pecaut \& Mamajek
2012, in prep.) using the RC spectrograph on the
SMARTS\footnote{http://www.astro.yale.edu/smarts/} 1.5-m telescope at
Cerro Tololo. Sco-Cen is the nearest OB association to the Sun
\citep[mean subgroup distances of $d$ $\simeq$ 118-145
pc;][]{deZeeuw99} and consists of three subgroups with ages of
$\sim$11-17 Myr \citep{Pecaut12, Preibisch08}.  The survey sample
consisted of $\sim$350 stars with optical/near-IR colors consistent
with having K/M spectral types, PPMX proper motions \citep{Roeser08}
consistent with membership to the three Sco-Cen subgroups, and X-ray
emission detected in the ROSAT All-Sky Survey \citep{Voges99,
  Voges00}.  The photometric and astrometric survey PPMX catalog
\citep{Roeser08} is complete down to $V$ $\simeq$ 12.8 magnitude with
typical astrometric accuracy of 2 mas yr$^{-1}$.  The survey sample
was cross-referenced with the stars with light curves in the the first
public data release (DR1) of the SuperWASP public
archive\footnote{http://www.wasp.le.ac.uk/public/} \citep{Butters10}.
SuperWASP is a photometric sky survey for detecting transiting
extrasolar planets with instruments in La Palma and in South Africa,
which have continuously monitored the sky since 2004 and the DR1
contains nearly 18 million light curves \citep{Pollacco06}.  Of our
$\sim 350$ Sco-Cen survey stars, at least 200 appear to be new bona
fide pre-MS stars, and SuperWASP light curves were available for 138
of them.

Among the SuperWASP DR1 data for the new Sco-Cen members,
we\footnote{The deep SuperWASP eclipse for this star was first noticed
  3 December 2010 by Pecaut \& Mamajek.} identified a star with a
remarkable light curve (PPMX J140747.9-394542 = GSC 7807-0004 = 1SWASP
J140747.93-394542.6 = 3UC 101-141675 = 2MASS J14074792-3945427;
hereafter ``J1407'').  The light curve is dominated by a
quasi-sinusoidal component with amplitude $\sim$0.1 mag in the WASP-V
photometric band, with periodicity of 3.21 days (consistent with
rotational modulation of starspots, typical for young active stars),
and a deep eclipse with maximum depth $\sim$3 mag between HJD 2454213
(23 Apr 2007) and HJD 2454227 (7 May 2007), with a complex pattern of
roughly symmetric dimming and brightening within $\pm$26 days of 29
Apr 2007 HJD 2454220.  The properties of this star are listed in Table
\ref{starprops} and the relevant optical/IR photometry is listed in
Table \ref{starphot}.  We discuss the eclipse further in
\S\ref{lightcurve}.

\begin{deluxetable}{lllllrlrcrrrr}
  \setlength{\tabcolsep}{0.03in} \tablewidth{0pt}
  \tablecaption{Properties of Star\label{starprops}} \tablehead{
    {(1)}&{(2)}&{(3)}\\
    {Property}&{Value}&{Ref}} \startdata
  $\alpha$(J2000) & 14:07:47.93 & 1\\
  $\delta$(J2000) & -39:45:42.7 & 1\\
  $\mu_{\alpha}$ & -25.4\,$\pm$\,1.4 \masyr\, & 1\\
  $\mu_{\delta}$ & -20.1\,$\pm$\,3.5 \masyr\, & 1\\
  Spec. type & K5 IV(e) Li & 2\\
  E(B-V) & 0.09 mag & 2\\
  A$_V$  & 0.32 mag & 2\\
  Dist  & 128\,$\pm$\,13 pc & 2\\
  EW(H$\alpha$) & 0.2 \AA (emis.) & 2\\
  EW(Li I $\lambda$6707) & 0.4 \AA (abs.) & 2\\
  T$_{eff}$ & 4500$^{+100}_{-200}$ K & 2\\
  \logl & -0.47\,$\pm$\,0.11 dex & 2\\
  R   & 0.96\,$\pm$\,0.15 R$_{\odot}$ & 2\\
  X-ray flux & 3.59 $\times$ 10$^{-2}$ ct s$^{-1}$ & 3\\
  HR1 & -0.04\,$\pm$\,0.42 & 3\\
  \loglxlbol & -3.4 & 2, 3\\
  L$_X$ & 10$^{29.8}$ \ergs & 2, 3\\
  P$_{rot}$ & 3.20 days & 2\\
  Age & $\sim$16 Myr & 2\\
  Mass & 0.9 M$_{\odot}$ & 2
\enddata
\tablecomments{References: (1) \citet{Zacharias10}, (2) this paper, (3) \citet{Voges00}.}
\end{deluxetable}

\begin{deluxetable}{lllllrlrcrrrr}
\setlength{\tabcolsep}{0.03in}
\tablewidth{0pt}
\tablecaption{Photometry of Star\label{starphot}}
\tablehead{
{(1)}&{(2)}&{(3)}&{(4)}\\
{Band}&{$\lambda_o$}&{Mag}&{Ref}}
\startdata               
V & 0.55 $\mu$m & 12.31\,$\pm$\,0.03 & 1\\
I & 0.79 $\mu$m & 10.92\,$\pm$\,0.03 & 2\\
J & 1.24 $\mu$m & 9.997\,$\pm$\,0.022 & 3\\
H & 1.66 $\mu$m & 9.425\,$\pm$\,0.023 & 3\\
K$_s$ & 2.16 $\mu$m & 9.257\,$\pm$\,0.020 & 3\\
W1 & 3.4 $\mu$m &  9.252\,$\pm$\,0.025 & 4\\
W2 & 4.6 $\mu$m & 9.276\,$\pm$\,0.021 & 4\\
W3 & 12 $\mu$m & 9.141\,$\pm$\,0.033 & 4\\
W4 & 22 $\mu$m & 8.907\,$\pm$\,0.388 & 4
\enddata
\tablecomments{(1) V-band is median of SuperWASP and ASAS measurements
  out of eclipse, given equal weight to both datasets, and the
  $\pm$0.03 mag is a systematic uncertainty.  The SuperWASP photometry
  was converted to Johnson V using factors by \citet{Bessell00} and
  assuming V$_{SuperWASP}$ = V$_{Tycho}$ \citep{Pollacco06}, (2) DENIS
  \citep{DENIS}, (3) 2MASS \citep{Cutri03}, (4) WISE first data
  release \citep{Wright10}.}
\end{deluxetable}

\begin{figure}[htb]
\epsscale{1.0}
\plotone{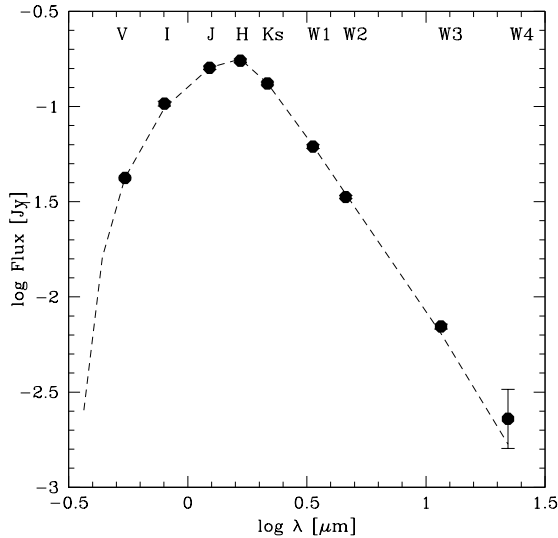}
\caption{Observed photometry for J1407 taken
from Table \ref{starphot} (filled circles) compared
to the spectral energy distribution for a lightly reddened
K5 dwarf with E(B-V) = 0.09 (red dashed line).
We have assumed that the K$_s$ minus WISE band (W1, W2, W3, W4) 
colors are zero.
\label{fig:sed}}
\end{figure}

\begin{figure}[htb]
\epsscale{1.0}
\plotone{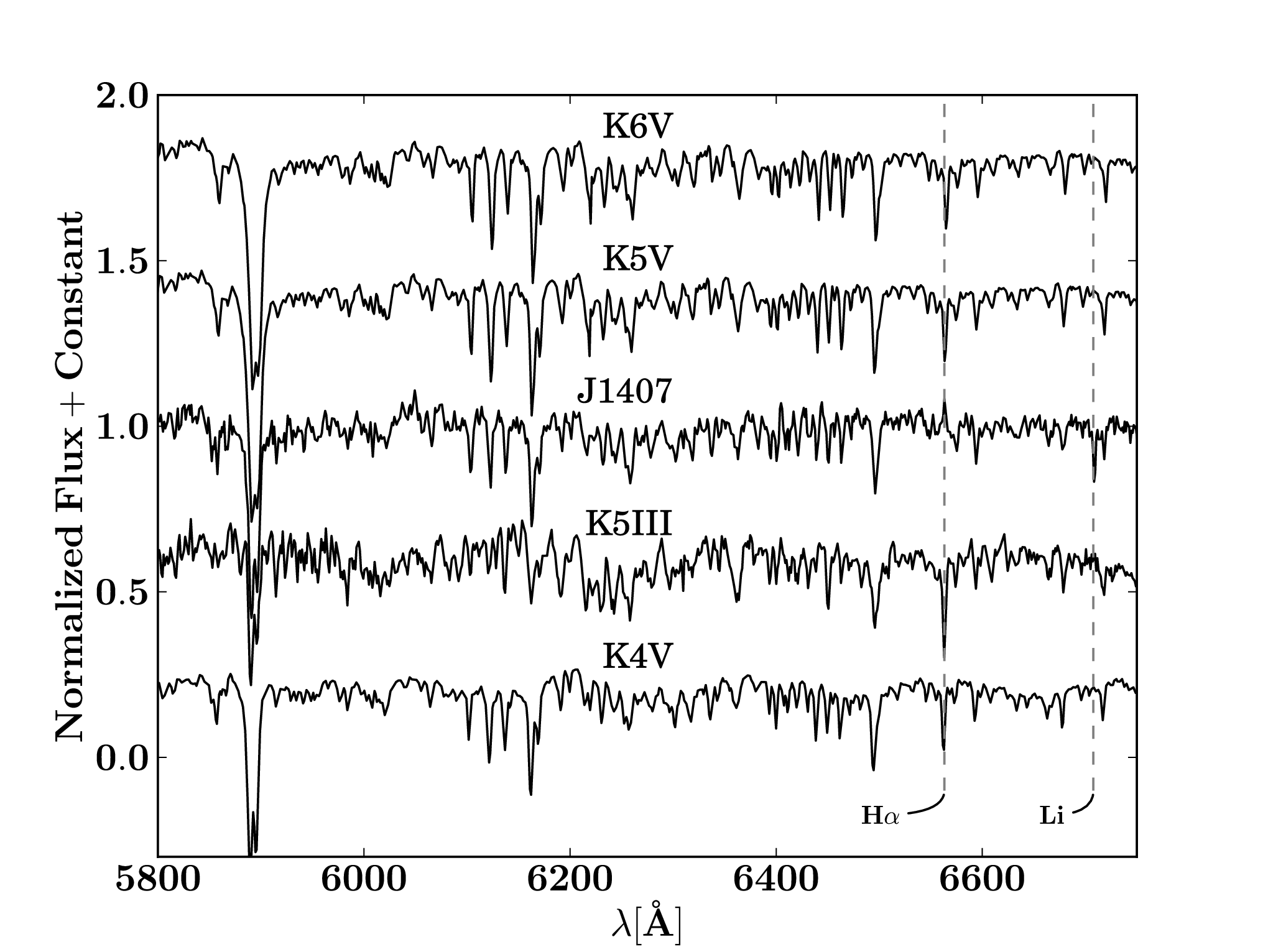}
\caption{Comparison of CTIO 1.5-m red optical spectrum of J1407 to CTIO
spectra of four spectral standard stars from \citet{Keenan89}: TW
PsA (K4V), N Vel (K5III), HD 36003 (K5V), and GJ 529 (K6Va). 
\label{fig:spec}}
\end{figure}

The spectral energy distribution for the star (plotted in Fig.
\ref{fig:sed}) is consistent with a lightly reddened K5 star (E(B-V) =
0.09, A$_V$ $\simeq$ 0.32 mag), with no evidence for infrared excess
\citep[via 2MASS and WISE preliminary release
photometry;][]{Skrutskie06, Wright10}.  Using a low-resolution red
spectrum taken in July 2009 with the SMARTS 1.5-m telescope RC
spectrograph, shown in Figure \ref{fig:spec}, we classify the star as
{\it K5 IV(e) Li}, i.e. a Li-rich K-star with negligible H$\alpha$
emission (0.2\AA\, equivalent width; i.e. ``filled-in''), and Na
doublet feature that is weaker than that for dwarfs, but not
consistent with a giant either \citep[so we adopt intermediate
luminosity class IV;][]{Keenan89}.\\

\begin{figure}[htb]
\epsscale{1.0}
\plotone{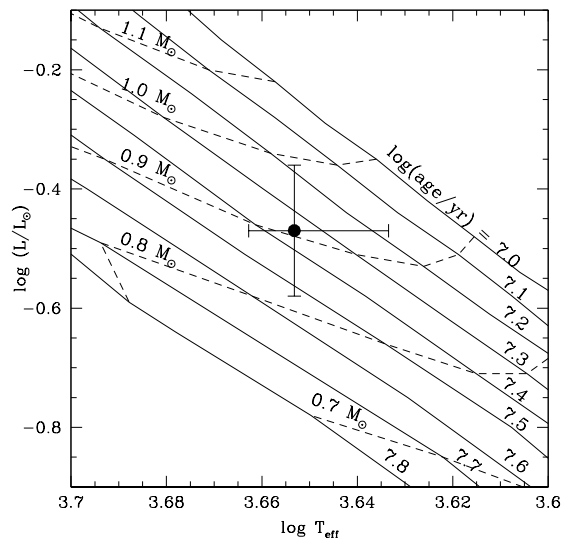}
\caption{HR diagram position for star J1407 with isochrones in
  log(age/yr) from \citet{Baraffe98} overlain.\label{fig:hrd}}
\end{figure}

The star is located in the vicinity of the Upper Centaurus-Lupus (UCL)
subgroup of the Sco OB2 association \citep{deZeeuw99}, and its proper
motion is statistically consistent with moving towards the UCL
convergent point (negligible peculiar velocity of 0.9\,$\pm$\,1.8
km/s), with a kinematic distance of 128\,$\pm$\,13 pc (similar to
other UCL members)\footnote{The distance and peculiar velocity were
  calculated following \citet{Mamajek05}, using the UCAC3 proper
  motion for J1407, and the updated estimate of the mean space motion
  for UCL from \citet{Chen11}: (U, V, W) = (-5.1$\pm$0.6,
  -19.7$\pm$0.4, -4.6$\pm$0.3) \kms.}.  Using this distance, we place
the star on the HR diagram, see Figure \ref{fig:hrd} ($\log T$, $\log
L/L_\odot$ = 3.66, -0.47).  Factoring in the $\pm$0.11 dex uncertainty
in \logl, dominated by the distance uncertainty, the isochronal ages
and their uncertainties are listed in Table \ref{tab:ages}.  Factoring
in previous age estimates for the UCL subgroup \citep[see summary
in][]{Mamajek02}, and new age estimates using the F-star MS turn-on
\citep{Pecaut11}, we estimate the mean age of UCL to be 16 Myr with a
$\pm$2 Myr (68\%CL) systematic uncertainty.  The isochronal age for
J1407 is consistent with this value, hence we adopt the mean UCL age
as the age for J1407. Three sets of evolutionary tracks predict
similar masses for J1407: $\sim$0.9 M$_{\odot}$ (listed in Table
\ref{tab:ages}).

The kinematics of the star, rotation period (as discussed below),
X-ray emission, and its preliminary HR diagram position, are mutually
self-consistent with the interpretation that this star is a nearby
(distance $\sim 130$ pc), $\sim 10^7$ year old, solar-mass pre-main
sequence (Pre-MS) star.

\begin{deluxetable}{lcl}
\setlength{\tabcolsep}{0.03in}
\tablewidth{0pt}
\tablecaption{Isochronal Age and Mass Estimates for J1407\label{tab:ages}}
\tablehead{
{(1)}&{(2)}&{(3)} \\
{Age}&{Mass} &{Models} \\
{Myr}&{\msun}&{...}}
\startdata               
 23$^{+14}_{-9}$  & 0.90\,$\pm$\,0.08 & 1\\
 27$^{+15}_{-10}$ & 0.89\,$\pm$\,0.07 & 2\\
 14$^{+11}_{-6}$  & 0.86\,$\pm$\,0.06 & 3  
\enddata
\tablecomments{References for the models are as follows: (1) \citet{Baraffe98}, (2) \citet{Siess00}, (3) \citet{DAntona97}.}
\end{deluxetable}

\subsection{Light Curves \label{lightcurve}}

\begin{figure}
\epsscale{1.0}
\plotone{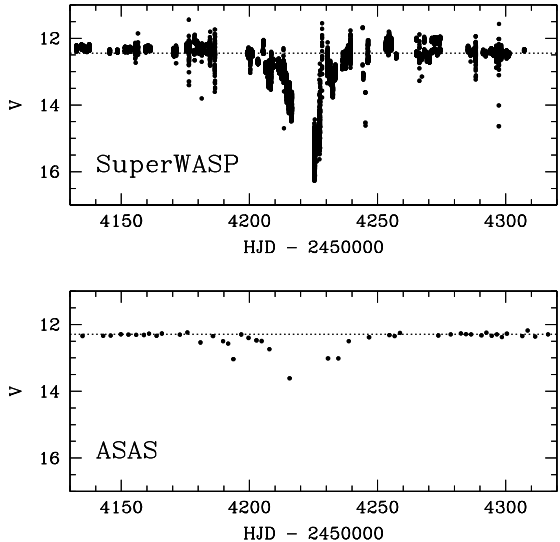}
\caption{Individual measurements of SuperWASP ({\it top}) and ASAS
  ({\it bottom}) V magnitudes for J1407 during early 2007.  Abscissa
  is Heliocentric Julian Date minus 2450000. 1 Jan 2007 midnight
  corresponds to HJD 2454101.5.  The eclipse was seen in both
  photometric data sets.  The eclipse was deep for about 14 days but
  is bookended by a gradual dimming covering a period of about $\pm$54
  days.  Long term median magnitudes outside of eclipse are plotted
  with dotted lines (V = 12.29 for ASAS, V = 12.45 for SuperWASP).
  The systematic difference is mostly due to SuperWASP-V being
  calibrated to the Tycho V$_T$ band \citep{Pollacco06}, whereas
  the ASAS is converted to the Johnson V system via 
  Hipparcos \citep{Pojmanski02}. 
  \label{fig:lightcurve}}
\end{figure}

\begin{figure}[htb]
\epsscale{1.0}
\plotone{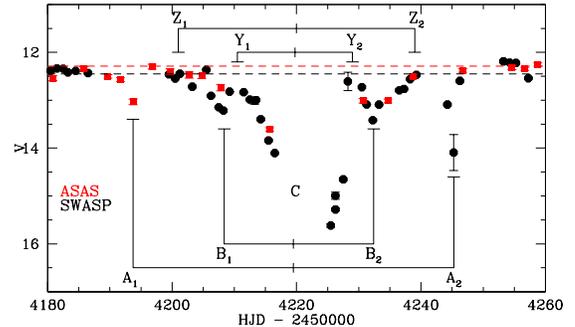}
\caption{SuperWASP and ASAS photometry for the April-May 2007 eclipse
  event(s). ASAS photometry is a single measurement each night,
  whereas the SuperWASP magnitudes are nightly median values (the
  standard error of the median is plotted). The four (two pairs)
  peripheral dips are labeled and matched to their partner. The midway
  date for both the ``A'' and ``B'' dips coincide within a day of HJD
  2454220 (29 Apr 2007). Dips A$_1$ and A$_2$ are $\sim$51.5 days
  apart, and dips B$_1$ and B$_2$ are $\sim$24 days apart.  Periods of
  low extinction are labelled Z$_1$ and Z$_2$ and Y$_1$ and Y$_2$.
  \label{fig:zoom}}
\end{figure}

\begin{figure}
\epsscale{}
\plotone{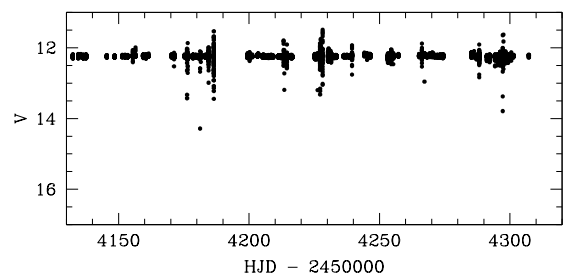}
\caption{
SuperWASP light curve for the comparison star
1SWASP J1407252.03-394415.1 (GSC 07807-00572), a star of similar
brightness (V = 12.24) to J1407 and situated 99''.6 away from it.
The time span covers the same range as the light curve for J1407 
in Fig. \ref{fig:zoom}. 7829 photometric data points with
median photometric error $\pm$0.018 mag are shown
and 68\% (95\%) are within $\pm$0.021 (0.077) mag of V = 12.245. 
Despite the appearance of some discrepent photometric points (with 
correspondingly large photometric errors), there is no evidence for 
any complex behavior similar to that seen for J1407. 
\label{fig:comp}}
\end{figure}

The V-band light curve for J1407 during the year of 2007 from the All
Sky Automated Survey (ASAS; \citep{Pojmanski02}) and SuperWASP (Super
Wide Angle Search for Planets) surveys is shown in Figure
\ref{fig:lightcurve}.  The SuperWASP survey is an ultra-wide field
(over 300 sq. degrees) photometric survey is designed designed to
monitor stars between V $\sim$ 7 -- 15 mag to search for transiting
extrasolar planets \citep{Pollacco06}. The public data archive of
SuperWASP photometry is described in \citet{Butters10}.  The SuperWASP
DR1 photometry for J1407 contains photometry for approximately 29,000
epochs during 206 dates between HJD 2453860 (2006.34) and HJD 2455399
(2010.56), with median photometric precision of 0.023 mag.  The light
curve for J1407 from the SuperWASP data\footnote{Can be retrieved from
  \url{http://www.wasp.le.ac.uk/public/lc/index.php} with the
  identifier 1SWASP J140747.93-394542.6} is dominated by (1) a
sinusoidal component with amplitude $\sim $ 0.1 magnitude in the
WASP-V photometric band, with periodicity of 3.211 days
\citep[consistent with rotational modulation of starspots, typical for
young active stars, e.g.][]{Mamajek08}, and (2) a 14 day deep eclipse
of depth $\ga 3$ mag between date HJD 2454213 (23 Apr 2007) and HJD
2454227 (7 May 2007), bookended by a gradual dimmings and brightenings
to the median brightness (Fig. \ref{fig:lightcurve}).  The same light
curve is also shown in Figure \ref{fig:zoom} using median nightly
SuperWASP values. In Fig. \ref{fig:comp} we plot the SuperWASP light
curve for a comparison field star situated $\sim$100'' away from
J1407, and of similar brightness (plotted over the same time period as
J1407's eclipse and the same magnitude scale as in Fig.
\ref{fig:zoom}). There is no evidence for similar complex behavior
during the epoch of J1407's eclipse in the light curve for the
comparison star. We discuss various scenarios for explaining the
dimming of J1407 in \S\ref{explanations}.

\begin{figure}[htb]
\epsscale{1.0}
\plotone{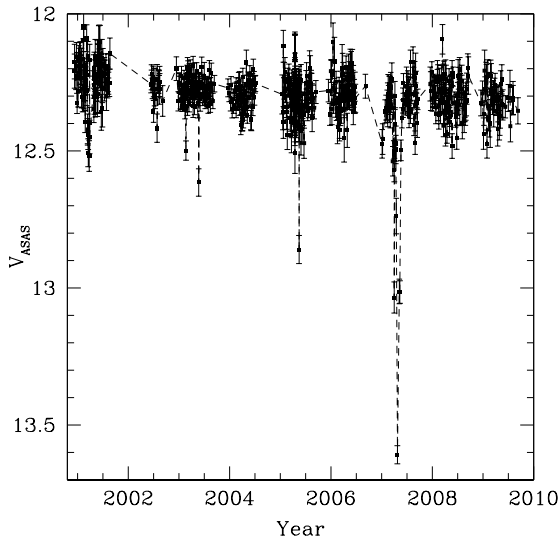}
\caption{ASAS light curve for J1407 between Feb 2001 and Sept 2009.
  An unconfirmed shallower eclipse might have occurred in early 2001. 
  Magnitudes are reported for a given night, and connected by a 
  thin dashed line
  \label{fig:asas}}
\end{figure}

The ASAS-3 archive also contained an extremely long time-baseline
light curve for J1407 (ASAS J140748-3945.7), with photometry provided
over 599 dates between Feb 2001 and Sep 2009 (more specifically, HJD
2451887 and HJD 2455088). The ASAS light curve is plotted with the
SuperWASP photometry during the eclipse in Fig. \ref{fig:lightcurve},
and the entire ASAS light curve for 2001-2009 is shown in Fig.
\ref{fig:asas}.  ASAS data also shows a minor, but sustained dip in
magnitudes between 2001.20 and 2001.24 ($\sim$14 days) of approximate
depth $\sim 0.2$ magnitudes (see Figure \ref{fig:asas}).  This
magnitude difference is only 2$\sigma$ above the night to night
dispersion however because the dimming lasted a couple of weeks, the
event stood out as unusual. If it was a true secondary eclipse, then
the period should be 12.24 years and the next secondary eclipse would
take place around 2013.46. High cadence photometry of J1407 in
mid-2013 should be able to test the idea that the 2001 event might
have been a secondary eclipse.

A series of nightly V-band images were taken of J1407 with the CTIO
1.3-m telescope in queue mode during the first half of 2011.  Three
consecutive 10 sec images were taken nightly during 106 nights between
7 Feb 2011 and 22 Jun 2011. Visual examination of the data, and
comparison of the brightness of J1407 to neighboring stars of similar
brightness shows no evidence of for deep ($>$0.5 mag) eclipses during
this period.

\subsection{Eclipse Substructure}

While the deepest part of the eclipse is not well sampled in Figures
\ref{fig:lightcurve} and \ref{fig:zoom} the eclipse of J1407 is
asymmetric.  \citet{Mikolajewski99} proposed that the asymmetry of EE
Cep's eclipses were due to the disk impact parameter with the line of
sight, and given the similarities of the central parts of their
eclipses, we suspect the same for J1407.  Nightly averages of
SuperWASP data (see Figure \ref{fig:zoom}) exhibit two pairs of
multi-day dips, labelled A$_1$ and A$_2$, separated by $\sim$24 days
and B$_1$ and B$_2$ separated by $\sim$51.5 days.  Between these dips,
there are periods that appear to be free of extinction lasting a few
days each, indicative that there may be large gaps in the disk.
\citet{Galan10} proposed that similar dips in the 2008/09 EE Cep
eclipse were due to gaps in a multi-ring disk.

\begin{figure}
\epsscale{1.1}
\plotone{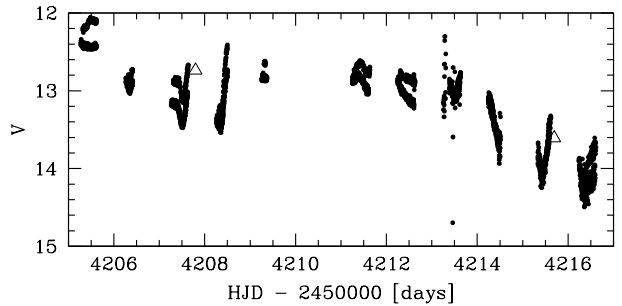}
\caption{SuperWASP ({\it dark filled circles}) and ASAS ({\it open
    triangles}) photometry during the portion of the eclipse
  immediately before the deepest minimum. During some nights the
  SuperWASP photometry shows (unphysical) jumps between two
  photometric levels at the tenths of magnitude level - a systematic
  effect seen in other SuperWASP studies \citep[e.g.][]{Norton11}.
\label{fig:zoom1}}
\end{figure}

\begin{figure}
\epsscale{1.1}
\plotone{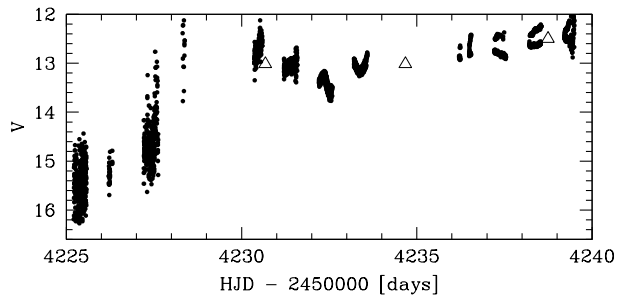}
\caption{SuperWASP and ASAS photometry for the 2nd part of the eclipse; 
same as for Fig. \ref{fig:zoom1}. 
\label{fig:zoom2}}
\end{figure}

There may also be substructure on timescales shorter than a day with
variations up to 1 mag on timescales shorter than a day.  Figures
\ref{fig:zoom1} and \ref{fig:zoom2} show detailed structure at the
beginning and end of the eclipse.  If this substructure is due to the
occulting body then it contains a remarkable wealth of structure.  We
have examined SuperWASP light curves of neighboring stars of similar
magnitude, as well as J1407 outside of eclipse, and seen no such
variations, implying that the hourly variations are due to the
eclipse. We explore hypotheses to explain the eclipses in Sec. 3.6,
and develop a model in Sec. 3.7.

\subsection{Rotation Period and X-ray Emission}

Young stars show variability on the timescale of days, induced by the
presence of starspots on the rotating surface. Variability at the
$\sim$0.1 mag level can be seen in the data, so we carried out a
search for periodicity in the star's light curve to determine the
star's rotational period. The ASAS photometry and SuperWASP photometry
are measured in slightly different bandpasses, and we measure this
magnitude difference by taking the median magnitude of the data in
each data set over the 2008 season, where there is no sign of long
term trends in the photometric light curve. We measure a systematic
offset of 0.143 magnitudes between ASAS and SWASP $V$-band photometry
(likely due to the SWASP photometry being calibrated to the Tycho
V$_T$ system), so we add an offset to the SWASP photometry to put it
on the ASAS $V$ system. For the SWASP data, we calculate the median
magnitude of each night and use this for the subsequent analysis.

\begin{figure}[htb]
\includegraphics[angle=-90,scale=0.6]{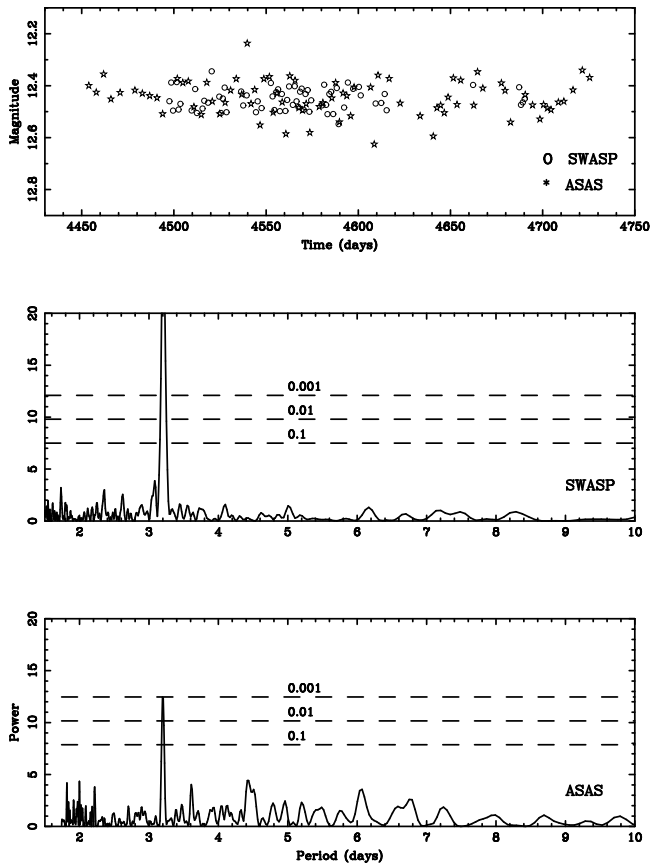}
\caption{{\it Top:} Photometry of the star over the 2008 season,
where the SuperWASP (SWASP) points are daily median values. 
{\it Middle:} Lomb Scargle periodogram for SuperWASP photometry,
and {\it Bottom:} Lomb-Scargle periodogram for the ASAS photometry.
  False alarm probabilities are indicated with horizontal dashed
  lines. A period of 3.20 days, presumably due to starspots and
stellar rotation, is detected strongly and independently in both
datasets. 
\label{fig:periodogram}}
\end{figure}

We take the photometry from SWASP and ASAS and perform a Lomb-Scargle
periodogram analysis on both data sets. False alarm probabilities
(FAPs) are estimated using the method described in \citet{Press92}.
The photometry over the 2008 season is shown in the top panel of
Figure \ref{fig:periodogram}, along with the Lomb-Scargle periodograms
of the SWASP and ASAS data in the lower panels. Both light curves show
a highly significant periodicity of 3.20 days, with FAPs of 10$^{-3}$
and 10$^{-6}$ for the ASAS and SWASP datasets respectively, with no
other detectable periods seen over the sampled period ranges.

The star has an X-ray counterpart in the ROSAT All-Sky Survey Faint
Source Catalog \citep[RASS-FSC;][]{Voges00}, with marginally detected
flux of $f_X$ = 0.0359\,$\pm$\,0.0148 ct s$^{-1}$, and hardness
ratioes of HR1 = -0.04\,$\pm$\,0.42 and HR2 = 0.06\,$\pm$\,0.62.
Using the energy conversion factor of \citet{Fleming95}, this
translates to an X-ray flux in the ROSAT band of $f_X$ = 2.9 $\times$
10$^{-13}$ erg s$^{-1}$ cm$^{-2}$, and using our previous distance and
bolometric luminosity estimates, \lx\, = 10$^{29.8}$ \ergs\, and
\loglxlbol\, $\simeq$ -3.4 dex.  A K5-type star with rotation period
of 3.20 days would be predicted to have soft X-ray emission around the
saturation level \citep[\loglxlbol\, $\simeq$
-3.2\,$\pm$\,0.3;][]{Pizzolato03}, perfectly consistent with the
observed ROSAT X-ray flux (\loglxlbol\, $\simeq$ -3.2), and consistent
with other Sco-Cen pre-MS stars \citep{Mamajek02}.

\subsection{Constraints on the Eclipse Period}

We detect a single deep eclipse in our data set, but there is the
possibility that another deep eclipse occurred during a period when
there was no photometric coverage. We determine if there could be
other eclipses that are undetected in the data set by choosing a trial
period for the eclipses and plotting the light curve modulo this
period. We count how many folded photometric points lie within the
deepest part of the known eclipse, and determine the mean magnitude
and standard deviation of these points.

\begin{figure}[htb]
\includegraphics[angle=-90,scale=0.6]{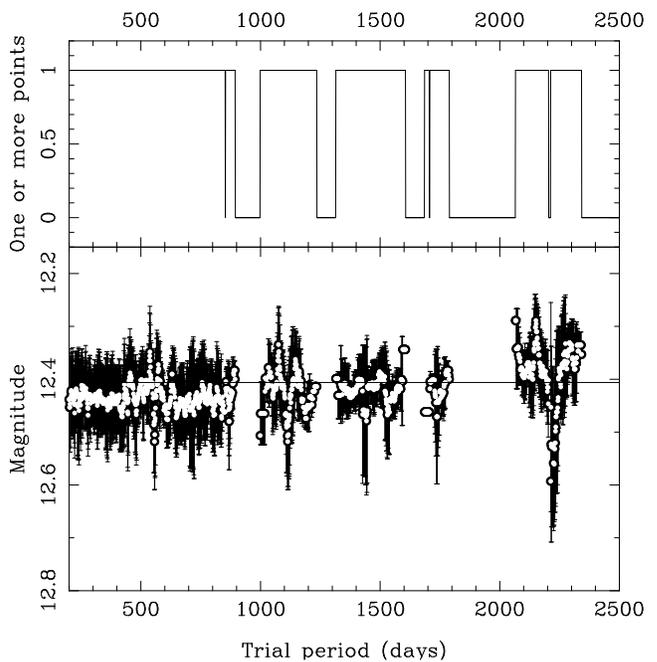}
\caption{Photometric coverage for periodic events at different trial
  periods. The upper panel shows the periods where we have one or more
  photometric points within an eclipse event, and the lower panel
  shows the mean magnitude in those particular cases.
  \label{fig:period}}
\end{figure}

We define a deep eclipse event as one where the stellar magnitude
fades by one magnitude or greater. For the known eclipse, we estimate
that this lasts for 15 days, and we set the trial periods P for values
starting at 200 to 2500 days in steps of one day. In Figure
\ref{fig:period} we show the results of our analysis. In the upper
panel we show that we have one or more photometric points for all
trial periods up to 850 days, and the lower panel shows the mean
magnitude of the photometric points at those periods where we have one
or more photometric points.  We conclude that there is no evidence for
any eclipse events with photometric coverage up to 2330 days (6.4
years), and that there are no eclipse events in any periods up to 850
days (2.3 years).  Approximately half of periods between 850 days and
2330 days are ruled out (Fig. \ref{fig:period}).

If the shallow depression seen in early 2001 (see Figure
\ref{fig:asas}) is an eclipse then the period is 6.12 years and just
about at the 2330 day limit.  EE Cep has a similar period and duration
and also exhibits variations in eclipse depth \citep{Galan10}.  Hence
we should consider the possibility that J1407 also exhibits large
variations in eclipse duration and depth.

\subsection{Other Explanations \label{explanations}}

The case for the primary being a pre-MS K star at d $\simeq$ 130 pc
seems to be secure. The primary exhibits: (1) rapid rotation (P = 3.2
days), (2) complimentary saturated X-ray emission consistent with the
rapid rotation (typical for young dwarf or pre-MS stars), (3) strong
Li consistent with other pre-MS Sco-Cen members, (4) proper motion
statistically consistent with Sco-Cen membership, and predicted
kinematic distance harmonious with pre-MS status, and lastly (5)
spectral appearance consistent with being dwarf or pre-MS, but not a
giant.

Besides the apparent spectral evidence, there are reasons to exclude
J1407 as a possible giant or supergiant.  With these observations of
the primary in mind, we briefly present and pass judgement on several
hypotheses regarding the agent responsible for J1407's unusual
eclipses.  If J1407 were a K5 giant with absolute magnitude similar to
the K5III standards N Vel and $\gamma$ Dra (\mv\, $\simeq$
-1.2)\footnote{Using Hipparcos V magnitudes \citep{ESA1997} and
  revised Hipparcos parallaxes \citep{vanLeeuwen07}, we estimate that
  the high quality K5 III spectral standards N Vel and $\gamma$ Dra
  \citep{Keenan89} have absolute magnitudes of \mv\, $\simeq$ -1.19
  and -1.14, respectively, and the K5 Ib standard $\sigma$ CMa has
  \mv\, $\simeq$ -4.3 \citep[assuming extinction \av\, $\simeq$ 0.12
  mag ;][]{Bobylev06}.}, then its apparent V magnitude would be
consistent with a distance of $\approx$4.3 kpc. J1407's total proper
motion (32 \masyr) would then imply a tangential velocity of
$\approx$670 \kms, i.e. faster than the local Galactic escape
velocity.  Photometrically, there is no hint of long-term periodicity
characteristic of red giants (i.e. Mira variability).  The situation
is worse if the star were a K5 supergiant. If it shared the absolute
magnitude of the K5 Ib standard $\sigma$ CMa (\mv\, $\simeq$ -4.3),
then J1407 would represent a young, massive star $\approx$18 kpc away,
$\approx$6 kpc above the Galactic disk midplane, with tangential
velocity of $\approx$2800 \kms. The presence of strong Li absorption,
X-ray emission, and a 3-day periodicity would also seem
extraordinarily unusual for a K5 giant or subgiant. Hence, we rule out
J1407 being an evolved, giant or supergiant late-K star.

We discuss some of the possible explanations for the observed eclipse,
and pass judgement on their plausibility.

\begin{itemize} 

\item {\it Eclipses by stellar or substellar companion alone}: No
  plausible stellar or substellar companion can be responsible for
  dimming the K5 pre-MS star J1407 by more than 3 magnitudes ($>$95\%
  dimming), and eclipses by such an object would not explain the
  irregular shape, depth, and duration of the eclipse.

\item {\it Is the ``primary'' a red giant that is eclipsing a fainter,
    bluer star?} The rare cases of eclipsing binaries that eclipse by
  $>$few magnitudes are usually cases of a red giant transiting a
  smaller, hotter dwarf star \citep[e.g. RV Aps; A2V+K4III; depth
  $\simeq$ 1.5 mag;][]{Khaliullin06} or symbiotic binaries \citep[e.g.
  AR Pav; depth $\simeq$ 6 mag, period $\simeq$ 604
  day;][]{Quiroga02}.  There is no hint from the spectrum of J1407
  that it could contain a giant star, a hot component, or constitute a
  symbiotic binary.  As stated before, we think we can safely exclude
  the hypothesis that the J1407 primary is an evolved late-K giant or
  supergiant.  Such a scenario would also not explain the eclipse
  structure in the weeks before and after the main deep eclipse.

\item {\it Could the obscuration be associated with a disk orbiting a
    compact stellar remnant?}  The system is too young to contain a
  neutron star or white dwarf.  Given the age of the system ($\sim$16
  Myr), any black hole would have had a progenitor mass of $>$14
  M$_{\odot}$ \citep{Bertelli09} and would have been an extremely
  large red supergiant and/or Wolf-Rayet star before its supernovae.
  A black hole would likely be a much stronger source of X-rays
  \citep[L$_X > 10^{32}$ \ergs;][]{Verbunt93} than observed (L$_X$
  $\simeq$ 10$^{30}$ \ergs) if it accreted from a disk. Also, the
  system's proper motion appears to be comoving with Sco-Cen within
  $\sim$1 \kms, so if there were a companion that supernovaed and
  removed a substantial amount of mass from the system, then why is
  J1407 not a runaway star? Hence it seems very implausible that the
  obscuration is associated with a disk orbiting any type of stellar
  remnant.

\item{\it Can a circumbinary or circumstellar disk about the star
    explain the obscuration?}  The pre-MS binary system KH 15D has
  exhibited photometric variations and eclipses over the past half
  century that are attributed to the effect of a precessing
  circumbinary disk \citep{Herbst10}.  The eclipses are 3.5 mag deep
  in all bands and last for a significant fraction (one third or 16
  days) of the orbital period (48 days).  The single star V718 Persei
  in the young cluster IC 348 also exhibits prolonged ($\sim 1$ year)
  eclipses of about 1 mag with a period of 4.7 yr.  Its eclipses are
  attributed to an inner edge of a circumstellar disk \citep{Grinin08,
    Grinin09}.  For both of these systems the eclipses last a
  significant fraction of the orbital period implying that the
  eclipsing object nearly fills the orbital plane.  V718 Per has a
  weak IR excess corresponding to a ``thin, low-mass disk''
  \citep{Grinin08}. KH 15D has negligible mid-IR excess (C.
  Hamilton-Drager, priv. comm.). J1407 lacks a near- or mid-infrared
  excess that would indicate a warm dust disk of substantial optical
  depth (and the lack of strong emission lines in the spectrum also
  indicates no evidence for an accretion disk).  Including the gradual
  dimming phase, the eclipse on J1407 lasted about 54 days.  As
  discussed in the previous section the period analysis suggests that
  the orbital period P $>$ 850 days (2.33 yr) so the ratio of eclipse
  duration to orbital period must be less than 0.06.  This is
  significantly lower than the ratios for either V718 Persei or KH 15D
  (0.2 \& 0.3, respectively) and suggests that a circumbinary or
  circumstellar disk about the K star cannot account for the eclipse.

\item{\it Could the eclipse be due to a circumstellar disk that
    occulted the star once due to the relative motions of the Sun and
    J1407?} We have not yet positively identified more than a single
  eclipse, so at present it is possible that the eclipse was a
  one-time occurrence.  As a UCL member, the tangential motion of
  J1407 on the sky is 32 \masyr\, or 20 \kms\, at its predicted
  distance of 128 pc. What if we interpret the obscuration as being
  due to a geometrically thin circumstellar disk orbiting J1407, with
  the optical depth changing due to the motion of the Sun relative to
  J1407? The eclipse depth would then suggest an optically thick
  mid-plane ($\Delta$mag $>$ 3.5 mag, or $\tau$ $>$ 3.2).  In a week,
  the Sun-J1407 line only sweeps 0.6 milliarcsecond due to their
  relative motion.  Assuming the disk to be of similar size to typical
  planetary orbits ($\sim$10 AU), this would translate to sweeping the
  disk in the z-direction approximately $\sim$4($r_{disk}$/10 AU) km
  per week.  A two week eclipse would correspond to a disk $\sim$10 km
  thick, suggestive of a remarkably thin disk with aspect ratio
  (height over radius) of $\sim$10$^{-8}$, similar to the rings of
  Saturn. Besides the thinness of the disk, there are other problems
  with this scenario: (1) it does not explain the nearly symmetric
  dimmings at $\pm$12 and $\pm$26 days from the inferred eclipse
  minimum, (2) the similarity with the periodic eclipsing object EE
  Cep would have to be coincidental, and (3) the future detection of
  another similar eclipse would obviously negate the idea.

\item {\it Could the eclipses be due to a circumstellar disk orbiting
    a star more massive than the K5 star?}  We can consider the
  possibility that the system is like $\epsilon$ Aurigae with the more
  massive object obscured by the eclipsing disk.  Hiding a dwarf star
  more massive than the K star seen would require an edge-on disk
  \citep[such as seen in images of HK Tau; ][]{Stapelfeldt98} but
  again there is no evidence for an infrared excess from J1407.  The
  $12\mu$m WISE flux corresponds to a flux $\lambda F_\lambda \sim 1.5
  \times 10^{-15}\,{\rm W~m}^{-2}$.  We have used the $12\mu$m flux
  because the signal to noise is significantly higher than that at
  22$\mu$m.  The error at 12$\mu$m is only 3\% of the flux so at best
  a disk could be emitting with a 12$\mu$m infrared flux of $\lambda
  F_\lambda \sim 5 \times 10^{-17}\,{\rm W~m}^{-2}$.  If only
  1/10$^{th}$ of the disk luminosity is emitted at 12$\mu$m then the
  total infrared flux of our source is at most $F_{IR} \lesssim 5
  \times 10^{-16}\,{\rm W~m}^{-2}$.  The solar luminosity at a
  distance of 128~pc (that estimated for J1407) corresponds to $2
  \times 10^{-12}\,{\rm W~m}^{-2}$.  Thus the total infrared
  luminosity is at most $L_{IR} \lesssim 2.5 \times 10^{-4} L_\odot$.
  It would be difficult to hide a main sequence star and remain below
  this luminosity, typical of debris disks, even if the disk were an
  edge-on transition disk with an inner hole.  It is more likely that
  the object hosting the occulting disk has a lower mass than the K
  star.  In this case the disk infrared luminosity could be consistent
  with this upper limit.

\end{itemize}

\subsection{A Preliminary Model \label{model}}

We now consider the possibility that the eclipse could be due to
occultation by a circumsecondary or circumplanetary dust disk with the
secondary object and its disk in orbit about the K5 primary star
(analogous to what has been proposed for EE Cep).  The period analysis
above suggests that the orbital period P $>$ 850 day (2.33 yr), which
for $m_1 = 0.9$ M$_{\odot}$ suggests an orbital radius of $>$1.7 AU
and an orbital velocity of $<$21.7 \kms.  The hypothetical disk
appears to produce some obscuration during a minimum of $\Delta t
\approx$ 54 days, and hence the obscuration occurs during a faction
$f_{eclipse} <$6.3\% of the period.  Using equation \ref{eqn:fe} for
the fraction of time spent in eclipse we find that this limit implies
that
\begin{equation}
m_2 \lesssim 21 M_J \xi^{-3},
\end{equation}
where we have used $m_1 = 0.9 M_\odot$ estimated for J1407.  If the
disk fills the Hill or tidal radius ($\xi\sim 1$) then the secondary
is likely to be a brown dwarf.  The secondary could have a higher mass
if the disk radius only partly fills the tidal radius estimated from
its semi-major axis.  If the secondary is in an eccentric orbit and
its disk is truncated tidally at pericenter then $\xi$ estimated from
the semi-major axis would be lower than 1.

If the dimming seen in 2001 corresponds to a secondary eclipse then
the fraction of the period spent in eclipse is $f_{eclipse} \sim
0.024$ corresponding to (using equation \ref{eqn:fe} and using an
eclipse time of 54 days and period of 12 years)
\begin{equation}
m_2 \sim  1.2 M_J \xi^{-3}.
\end{equation}
For $\xi \sim 0.2$ expected for circumplanetary disks this gives $m_2
\sim 0.1 M_\odot$ and an M dwarf.  Both this mass estimate and the
previous one estimated from the 850 day period limit suggest that the
mass of the companion must be low and in the brown dwarf or low mass M
star regime.  The longer the period, the lower $f_{eclipse}$, and the
lower the estimated mass of the secondary.  However if the period is
longer then the separation between primary and secondary would be
larger making it easier to resolve using a high angular resolution
imaging system.  For example, a separation of 10 AU at a distance of
130 pc corresponds to 77 mas, resolvable perhaps with aperture-masking
interferometry on large telescopes \citep[e.g.][]{Ireland08}.

The period and length of the eclipse can be used to estimate the disk
radius.  Inverting equation \ref{eqn:days}
\begin{eqnarray}
r_d &\sim & \left({t_{eclipse} \over P }\right) \left( {P \over 2 \pi}\right)^{2\over 3} \left[  G (m_1 + m_2) \right]^{1\over 3} \\
&\sim& 0.08 AU \left( {f_{eclipse} \over 0.06}\right) \left( { P \over 2.3 {\rm yr} }\right)^{2\over 3} \left({ m_1 + m_2 \over 0.9 M_\odot} \right)^{1\over 3}. \nonumber
\end{eqnarray}
As expected, only large objects are capable of causing such a long
eclipse.  The disk radius is only weakly dependent on the period
(proportional to $P^{-1/3}$) and would be smaller for a more distant
companion.

Regions during the eclipse with little dimming could be interpreted in
terms of gaps in the disk (as by \citealt{Galan10} for EE Cep).  The
gaps (regions labelled Z$_1$, Z$_2$, Y$_1$ and Y$_2$ on Figure
\ref{fig:zoom}) each last a few days.  The ratio of the width of these
gaps to one half of the eclipse time is approximately 3/26 or 0.11
suggesting that the disk must have an aspect ratio $h/r$ smaller than
this ratio.  If we assume that the gaps are twice the Hill radius of
an objected embedded in the disk then the ratio of the gap to total
eclipse time gives an upper limit on the ratio of the third power of
the ratio of the satellites to secondary mass.
\begin{equation}
{t_{gap} \over t_{eclipse}} \lesssim \left( {   m_s \over 3 m_2} \right)^{1/3} 
\end{equation}
where $m_s$ is the mass of the gap opening satellite.  For
$t_{gap}/t_{eclipse} \sim 0.06$ this corresponds to $m_s / m_2
\lesssim 10^{-3}$.  If the secondary has a mass of 1$M_J$ then the
satellites in the disk could have mass lower than that of Earth, and
if the secondary is an M star of mass $0.1 M_\odot$ then the gap
opening satellites could have mass lower than Saturn.

Photometric variations on daily timescales (e.g., see Figures
\ref{fig:zoom1}, \ref{fig:zoom2}) suggest that there are hourly
variations in the eclipse depth.  The ratio of a few hours (or a
quarter day) to the half eclipse length is about 0.01 suggesting that
the disk has an extremely low aspect ratio of $h/r \lesssim 0.01$.
{\it If so then the disk could not be a gaseous disk but must be a
  planetesimal disk or a ring system}.

We are currently attempting to model the eclipse light curve in terms
of an optically thick inner disk which caused the primary deep eclipse
and a system of rings of lower optical depth which caused the smaller
dips in the weeks before and after the primary eclipse (Scott,
Moolekamp, \& Mamajek, in prep.). We have had success modeling the
primary eclipse, however the peripheral dips have been more difficult
to model.  Light from the primary star is modeled as a collection of
spherically symmetric points \citep[][]{Wilson71} and a secondary star
or planet modeled as a sphere (we assume the tidal deformations of
both objects to be negligible because the length of the eclipse and
minimum period implies a large disk and distance to the primary). Limb
darkening was calculated following \citet{vanHamme93}.  The disk and
rings are assumed to be thin debris disks of dust with uniform density
and opacity.  The input parameters were the mass of the primary, the
orientation of the debris disk, the orientation, period and
eccentricity of the orbit, the inner and outer radius of each ring,
and the opacity of each ring.  Physical parameters as a function of
age and mass for the primary were taken from \citet{Baraffe98} and for
the low-mass companion from \citet{Baraffe02}.

\begin{figure}[htb]
\includegraphics[width=3.4in]{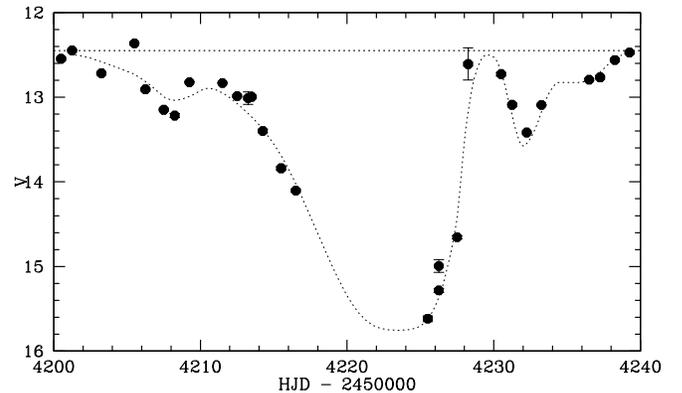}
\caption{A simple, non-unique model attempting to fit the gross
  features of the nightly mean SuperWASP photometry for J1407. This
  model contains around an object orbiting J1407 girded by a thick
  inner disk, a gap, and two rings with a small gap between them. The
  thick inner disk is used to model the deep eclipse feature ``C'' in
  Fig. 5, and the two rings are modeled to fit the features ``B$_1$''
  and ``B$_2$'' in the same figure.  The companion object has a test
  orbital inclination of 89$^{\circ}$.955, axial tilt with respect to
  the orbital plane of 13$^{\circ}$, orbital period 9862 days (a = 8.7
  AU), and radius R$_{c}$ = 1.46 R$_{Jup}$.  The model contains a 
  thick inner disk with $\tau_{\perp}$ = 0.5 and outer radius 76
  R$_{c}$, a first ``ring'' with optical depth $\tau_{\perp}$ = 0.2 
  between
  106 and 127 R$_{c}$, and a second ``ring'' with optical depth
  $\tau_{\perp}$ = 0.05 between 128 and 163 R$_{c}$. Yet another
  outer ring is needed to fit dips before and after 
  the time range plotted. 
  \label{fig:fit}}
\end{figure}

\begin{figure}[htb]
\includegraphics[width=3.4in]{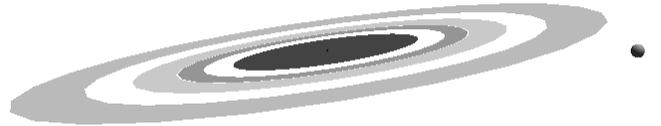}
\caption{Diagram of J1407's dust disk model whose corresponding light
  curve is plotted in Fig. \ref{fig:fit}. The K5 star J1407 is plotted
  to scale on the right (radius = 0.96 R$_{\odot}$). The thick ($\tau$
  = 3) inner disk is needed to produce feature ``C'' in Fig. 5, and
  the tilt is largely responsible for the asymmetric eclipse. The two
  annuli of lower optical depth ($\tau_{perp}$ = 0.2 and 0.05) are
  used to fit features B$_1$ and B$_2$ in the light curve. The gap
  between the thick inner disk and inner ring is needed to model
  features Y$_1$ and Y$_2$ in Fig. 5.  The outer ring hypothesized to
  explain features A$_1$ and A$_2$ in the light curve in Fig. 5 is not
  included in this diagram.  The intra-night SuperWASP light curves
  are suggestive of much more substructure within the ``rings'' than
  represented here.
  \label{fig:cartoon}}
\end{figure}

A preliminary good fit (but by no means unique) to the daily-averaged
light curve for J1407 is shown in Fig.  \ref{fig:fit}, and the model
parameters are listed in the caption and summarized in Table
\ref{ringparams}.  A diagram showing the geometry of this preliminary
good fit is shown in Fig. \ref{fig:cartoon}.  Besides a thick inner
disk needed to explain the deep eclipse labelled C in Fig. 5, our toy
model includes two ``rings'' of different optical depth for explaining
features B$_1$ and B$_2$ in the light curve.  An additional optically
thin outer ring is needed to explain features A$_1$ and A$_2$ in the
light curve. This outer ring is not included in this model, but has
the following approximate parameters: r$_{in}$ $\simeq$ 200 R$_c$
(where R$_c$ is the companion radius, for this model assumed to be
1.46 R$_{Jup}$ $\simeq$ 104,000 km), r$_{out}$ $\simeq$ 250 R$_c$,
$\tau_{\perp}$ $\simeq$ 0.09.  Another gap is needed to explain maxima
Z$_1$ and Z$_2$ in Fig. 5, with inner and outer radii of approximately
$\sim$163 R$_c$ and $\sim$200 R$_c$.

Approximately how much dust could be responsible for the eclipses that
we are seeing?  The densest of Saturn's named rings - the B ring - has
a radially-averaged optical depth of order unity and surface density
of $\sim$50 g\,cm$^{-2}$ \citep{Zebker85, dePater01, Tiscareno12},
implying an approximate opacity of $\kappa$ $\sim$ 0.02
cm$^{2}$\,g$^{-1}$.  Adopting this mass opacity for J1407's dust
annuli, the thick inner disk would have a mass of $\sim$0.8 \mmoon,
the two rings shown would have masses of $\sim$0.2 and $\sim$0.1
\mmoon, and the outermost ring would contain $\sim$0.5 \mmoon\, of
dust mass.  Our toy model assumes that the companion and its disk
system is situated 8.7 AU from J1407, however this is only
definitively constrained to be $>$1.7 AU given the time-series
photometry available.  So any estimates of the physical scale and mass
of the disk system will obviously scale with the companion's orbital
separation from J1407.  For $\kappa$ = 0.02 cm$^{2}$\,g$^{-1}$, the
total ring mass ranges from $\sim$3.6 \mmoon\, for $P$ = 2.33 yr, to
$\sim$0.8 \mmoon\, for $P$ = 200 yr. Over this same range, the outer
edge of the outermost ring scales from 60 million km (0.4 AU) for $P$
= 2.33 yr, down to 14 million km (0.09 AU) for $P$ = 200 yr.  We are
in the process of improving the code, attempting to fit the
intra-night light curves, and making predictions of brightness of the
companion's disk in the infrared, in order to give further constraints
on the size and orientation of the debris system, and the physical
parameters of the gaps and dust rings.

\begin{deluxetable}{lccllc}
  \setlength{\tabcolsep}{0.03in} \tablewidth{0pt}
  \tablecaption{Ring Model Parameters\label{ringparams}} 
\tablehead{
{(1)} &{(2)}     &{(3)}      &{(4)}   &{(5)}            &{(6)} \\
{Ring}&{R$_{in}$}&{R$_{out}$}&{$\tau$}&{$\tau_{\perp}$} & {Mass}\\
{}    &{(R$_c$)} &{(R$_c$)}  &{}      &{}               & {M$_{Moon}$}}
\startdata
 ``Rochester'' & 1?& 81 & 3.0 & 0.5  & 0.8\\
 ``Sutherland''&106&127 & 1.0 & 0.2  & 0.2\\
 ``Campanas''  &128&163 & 0.3 & 0.05 & 0.1\\
 ``Tololo''    &200&255 & 0.5 & 0.09 & 0.5\\
\enddata
\tablecomments{$R_{in}$ is inner radius, $R_{out}$ is outer radius,
  $\tau$ is optical depth through ring along line of sight between
  observer and the primary star, $\tau_{\perp}$ is the optical depth
  through the ring perpendicular to the ring place. Ring sizes are
  parameterized by companion radius, assumed to be 1.46 $R_{Jup}$
  (104,000 km).  This model assumes a companion orbital period of 27
  years ($a$ = 9.7 AU) and ring opacity of $\kappa$ = 0.02
  cm$^{2}$\,g$^{-1}$.  To scale the radii for different assumed
  orbital periods, multiply them by factor ($P$/27 yr)$^{-1/3}$ (where
  $P$ $>$ 2.33 yr). To scale the ring masses, multiply by ($P$/27
  yr)$^{-2/3}$($\kappa$/0.02 cm$^{2}$\,g$^{-1}$)$^{-1}$. Ring
  nicknames come from locations where observations were taken or
  analysis carried out for this study (The SuperWASP observations were
  taken using the SuperWASP-South observatory located at Sutherland,
  South Africa. The ASAS survey was carried out at Las Campanas
  Observatory, Chile.  Spectra of the host star were taken with the
  SMARTS 1.5-m on Cerro Tololo. Discovery and analysis of the system
  took place at University of Rochester).}
\end{deluxetable}

\section{The Probability of Seeing Eclipses by Circumsecondary or
  Circumplanetary Disks in a Sample of Young Stars}

We first estimate the probability of a circumplanetary disk eclipse
using probability distributions for giant planets estimated from
radial velocity surveys.  We then estimate the probability of a
circumsecondary disk eclipse based on surveys of young binary stars.

\subsection{Probability of seeing eclipses by circumplanetary disks}

To estimate the probability of detecting a circumplanetary disk
eclipse we must consider the number of stars that host gas giant
planets.  The period and mass distribution of gas giants estimated
from Doppler radial velocity surveys is
\begin{equation}
dN = C M^{-3.1\pm0.2} P^{0.26\pm0.1}  d\log M d\log P
\end{equation}
where the normalization constant $C$ is such that the fraction of FGK
stars with a planet in the mass range 0.3 - 10 $M_J$ (where $M_J$ is a
Jupiter mass) and period range 2-2000 days is 10.5\%
\citep{Cumming08}.  In units corresponding to measuring planet masses
in Jupiter masses and orbital periods in days, the value of the
normalization is $C =1.4 \times 10^{-3}$.  Integrating over the masses
we find $dN = C' P^{0.26} d\log P$ with $C' = 4.5 \times 10^{-3}$.
The distribution gives a probability of a FGK star hosting a gas giant
between 1-5 AU (365-1825 days), interior to this (2 to 365 days), and
from 5-20 AU, in each case of about $f_g \sim 0.05$ or 5\%.

We consider a sample of stars restricted so that they have already
depleted circumstellar disks (i.e. are post-accretion pre-MS stars)
but are young enough that they could host circumplanetary disks of
sufficient optical depth to produce detectable eclipses.  One could
choose a sample based on eliminating stars with evidence of accretion
and selecting for age based on chromospheric activity, cluster or
association membership \cite[e.g.][]{Mamajek02, Mamajek08}.  Typical
subgroups of OB associations have $\sim$10$^3$ stars, and ages of
$\sim$3-20 Myr \citep[e.g.][]{deZeeuw99, Briceno07}, during which the
majority of stars have just recently ceased accreting from
circumstellar disks \citep[e.g.][]{Mamajek09}. Light curves of
post-accretion $\sim$1 M$_{\odot}$ pre-MS stars in the nearest OB
associations (e.g. Sco-Cen, Ori OB1, etc.)  can be searched in
existing SuperWASP and ASAS datasets, and indeed such an effort is
currently underway by our group.

If each star in the sample is observed a single time the fraction that
would be observed in eclipse we estimate as $f_1 \sim f_{orient}
f_{eclipse} f_g$ where we multiply the number of systems with gas
giants, $f_g$, by the fraction of orbit spent in eclipse,
$f_{eclipse}$ (equation \ref{eqn:fe}), and fraction of orientations
capable of giving eclipse $f_{orient}$ (equation \ref{eqn:fo}).  For
our three ranges of semi-major axis radii
\begin{eqnarray}
  f_1 &\sim& \xi^2 \mu^{2/3} 3^{-2/3} \pi^{-1} \bar y f_g \\
  &\sim& 10^{-5.8} 
  \left({\xi \over 0.2} \right)^2 \left({m_p \over m_J}\right)^{2\over 3} \left({M_* \over M_\odot}\right)^{2\over 3} 
  \left({{\bar y} \over 0.5} \right) \left({f_g \over 0.05}\right) \nonumber
\end{eqnarray}
This probability is very low and implies that high-cadence monitoring
of many stars is required to detect circumplanetary disk eclipses.

We now consider the same sample of stars but continuously monitor them
throughout the planet's orbital period.  The fraction (for each range
of semi-major axis) that would exhibit an eclipse of a circumplanetary
disk would be $f_c \sim f_{orient} f_g$,
\begin{eqnarray}
  f_c &\sim&  10^{-3.7} \left({\xi \over 0.2} \right) \left({m_p \over m_J}\right)^{1\over 3} \left({M_* \over M_\odot}\right)^{1\over 3} 
  \left({{\bar y} \over 0.3} \right) \left({f_g \over 0.05}\right).
\end{eqnarray}
Restricting our study to planets between 1-5 AU (with $f_g \sim 0.05$)
and monitoring them for 10 years (approximately the orbital time at 5
AU for $\sim$1 \msun) the above fraction $f_c$ suggests that if we
monitor 10$^4$ stars $10^7$ years old then $\sim$2 of them should
exhibit circumplanetary disk eclipses.
   
A system with a recurring circumplanetary eclipse would make it
possible to study eclipses in depth, so discovery of systems with
short orbital periods are important.  For example, circumplanetary
disk substructure could be inverse modeled by observed high cadence
light curves during eclipse.  Unfortunately due to the smaller Hill
radius size closer to the star, we expect that the lifetime would be
short for circumplanetary disk in a smaller orbit about the star.  One
could search for circumplanetary disk eclipses in systems that have
not yet lost their outer circumstellar disks (for example systems such
as $\beta$ Pic or DM Tau).  However a circumstellar disk is a large
object and the probability that a circumplanetary disk occults the
star but the circumstellar disk does not occult the star is probably
even more negligible.
 
\subsection{Probability of Seeing Eclipses by Circumsecondary Disks}
 
We consider a 10 year photometric survey of a sample of weak-lined T
Tauri stars.  The fraction that would exhibit an eclipse by a disk
would be the fraction that are binaries ($\sim 0.5$;
\citealt{Duchene07, Kraus11}) times the fraction that have binary
periods less than 10 years, times the fraction that have a secondary
with an optically thick disk which could produce a significant
eclipse, times the probability that such systems are oriented in a way
giving eclipses (given by equation
\ref{eqn:fo}).
 
For young binary systems in star-forming regions the estimated
fraction of mixed systems with primary a weak lined T Tauri star and
the secondary a classical T Tauri star is not low, and could be as
large as $\sim 1/3$ \citep{Monin07}.  The fraction of young binary
systems that have a weak lined T Tauri primary and a secondary with a
passive non-accreting disk is lower; the survey described by
\citet{Monin07} contains only 1 out of about 80 binaries.  A tidally
truncated disk around a low mass secondary is expected to have a
shorter accretion lifetime than the primary's disk \citep{Armitage99},
though its planet formation timescale could be longer.  Thus a high
mass ratio binary (with mass ratio of order $q \sim 0.1$) with a
classical T Tauri phase primary and a passive disk about the secondary
should be relatively rare.  A mid-IR survey of about 65 binary young
stars finds that about 10\%\, contain passive dust disks
\citep{McCabe06} and for one of these the disk is hosted by the
secondary.  Thus of the binaries studied by \citet{Monin07} and
\citet{McCabe06} we can crudely estimate that 1/100 could be like
J1407 with a weak-line T Tauri primary and a low mass secondary with a
passive disk (however this fraction should diminish as one gets to
older pre-MS stars with ages of $>$10$^7$ yr).  As described by
\citet{Prato10}, binary systems with primaries ``...classified as
weak-lined T Tauris, unresolved, might also harbor truncated disks
around the secondary stars. Such small structures could go undetected
as the result of dilution from a relatively bright primary.
Circumstellar disks with central holes that show excesses in the
mid-infrared but not in the near-infrared, and which do not show
signatures of accretion, may be present but are effectively
undetectable.''  J1407 may be in this class and an example of a
relatively rare young binary with weak-lined T Tauri primary and low
mass secondary hosting a passive disk.

The number of binaries is flat in log period or semi-major axis space
\citep{Halbwachs03, Kraus11} and young binaries are similar to field
population in this respect \citep{Duchene07}.  Based on this
distribution about 1/3 of all binaries have periods less than 10 years
(dividing the semi-major axis ranges into three ranges 1-10 yr, 10-100
yr, and 100-1000 yr).  Thus the number of weak-lined T Tauri stars
that are binaries with periods less than 10 years and contain low mass
secondaries with passive disks would be of order 1/600.  Using a mass
ratio of 0.1, $\bar y=0.3$ and $\xi = 1$ we estimate the fraction
oriented such that they can give eclipses $f_{orient} \sim 0.1$.
Altogether a 10 year survey of weak-lined T Tauri stars might have a
probability of detecting an eclipse of order 1/6000 giving a similar
probability to that estimated above for circumplanetary disk eclipses.
An estimate folding the mass distribution and lifetime distributions
would improve this estimate.  However this is difficult to formulate
as binary identifications are not complete at mass ratios less than
0.1 \citep[e.g., ][]{Kraus11} and the number of secondaries hosting
passive disks in the $10^7$ year old age range is not well
characterized. There are hints that the protoplanetary disk fraction
decay timescale is systematically longer for lower-mass stars and
brown dwarfs compared to Sun-like stars and massive stars
\citep{Mamajek09}, however further observations will be useful in
constraining these findings for components of binary systems.
 
\section{Conclusion}

In this study we have estimated the probability that a system hosting
a gaseous circumplanetary or circumsecondary disk about a planet could
occult a star. The existence of circumplanetary disks after the
dissipation of the protosolar (circumstellar) nebular disk has been
postulated from formation scenarios for the Galilean satellites
\citep{Canup02, Magni04, Ward10}.  Because such a disk would be large
the probability that a system hosting one is oriented in such a way
that it can occult the star is tiny, but not zero.  Because the
lifetime of a circumplanetary disk could be longer for planets at
large semi-major axis, light detected from outer exoplanets such
Fomalhaut b may arise from such a disk.  We estimate that eclipses
from the thick inner circumplanetary disks that spawn regular
satellite systems around gas giants may last for days, however tenuous
outer disks of lower optical depths to larger fractions of the Hill
radius could persist for weeks, depending upon the planet's mass and
semi-major axis.  We estimate that a survey monitoring 10$^4$ stars
that are approximately $10^7$ years old for 10 years would likely
yield at least a few circumplanetary and circumsecondary disk eclipse
candidates. The 8.4-m Large Synoptic Survey Telescope
\citep[LSST;][]{Ivezic08} will photometrically monitor many thousands
of pre-MS stars in catalogued OB associations, young star clusters,
and star-forming regions during its scans of the Galactic plane
region. The LSST survey design should enable estimation of sub-mas/yr
proper motions which will allow kinematic membership assignments of
newly discovered pre-MS stars to young clusters and associations of
determinable distance and age. Optical spectroscopic follow-up will be
necessary to confirmation of youth (via Li absorption, c.f. our study
of J1407) and estimation of reddening for confirming that the star has
luminosity and isochronal age consistent with the other
cluster/association members. Even given LSST's proposed field revisit
time ($\sim$3 days, but twice per night), a J1407-like eclipse would
have been easily detected. Higher cadence follow-up imaging from
smaller dedicated telescopes will be necessary for detailed
characterization of the eclipse light curves, but LSST monitoring of
young stellar groups should yield some candidate disk eclipse objects.

In a survey of a few hundred $10^7$ year old stars we have discovered
a deep long eclipse in 2007 on the pre-main sequence K dwarf star
J1407.  This star was selected to be in the appropriate age range and
could host either a circumplanetary disk or a lower mass secondary
star with a disk.  The lack of infrared emission suggests that the
mass of the unseen object is much lower than that of the solar-mass K
star.  Limits for the period and the observed eclipse time suggest
that the unseen object hosting the disk is low mass, perhaps in the
substellar regime.  Substructure in the eclipse suggests that the disk
is thin $h/r\lesssim 0.01$ and has gaps that may contain satellites
with mass $\sim$$10^{-3}$ times that of the secondary.  Follow up with
a radial velocity study is important as radial velocity measurements
could put limits both on the period and mass of the secondary.

The complex eclipse of J1407 that took place in 2007 is slightly
asymmetric and contains significant substructure, similar to the
eclipses of the Be star EE Cep that have been interpreted in terms of
an occulting planetary system \citep{Galan10}.  As has been proposed
for this system, the asymmetry of the eclipse could be due to the
impact parameter of the disk with respect to the line of sight
\citep{Mikolajewski99}.  Variations in eclipse depth in this system
are attributed to a possible third companion that tilts the orbital
plane of the eclipsing system \citep{Torres00}.  Identifying a second
eclipse for J1407 will allow measurement of the eclipse period, and
planning of observing campaigns similar to those launched for EE Cep
and $\epsilon$ Aurigae.

Constraints from the gas giant satellite systems in our own solar
system suggest that their circumplanetary disk structures could have
produced quite complex eclipses if seen in transit, with dense inner
regions, gaps where satellite formation is taking place, and low
density disks possibly extending to large fractions of the Hill
radius.  Such eclipses seen among young stars may provide remarkable
laboratories for testing satellite and planet formation scenarios.
Regardless of the nature of the disked companion of J1407 (low-mass
star, brown dwarf, or gas giant planet), detailed observations of
future eclipses should provide useful constraints on either
circumsecondary or circumplanetary disk structure, and the early
evolution of planets and/or satellites.

\acknowledgements

We have used data from the WASP public archive in this research. The
WASP consortium comprises of the University of Cambridge, Keele
University, University of Leicester, The Open University, The Queen's
University Belfast, St. Andrews University and the Isaac Newton Group.
Funding for WASP comes from the consortium universities and from the
UK's Science and Technology Facilities Council. The star exhibiting
the unusual eclipse was discovered in a spectroscopic survey using the
SMARTS 1.5-m telescope, and the survey and support for EM and MP were
funded by NSF award AST-1008908.  EM, MP, FM, and ES acknowledge
support from the University of Rochester College of Arts and Sciences.
AQ acknowledges support through NSF award AST-0907841.  This research
has made use of the NASA/IPAC Infrared Science Archive, which is
operated by the Jet Propulsion Laboratory, California Institute of
Technology, under contract with the National Aeronautics and Space
Administration.  EM also thanks David James, John Subasavage, Andrei
Tokovinin, Warren Brown, and Catrina Hamilton-Drager for discussions,
and Fred Walter for scheduling queue observations of the star and
standards on the SMARTS 1.5-m.

{\it Facilities:}
\facility{SuperWASP},
\facility{ASAS},
\facility{CTIO:1.5m},
\facility{CTIO:1.3m}

\end{document}